# RKappa: Software for Analyzing Rule-Based Models

Anatoly Sorokin [1], Oksana Sorokina [2] and J. Douglas Armstrong [2]

NB: this manuscript was prepared for the Methods in Molecular Biology series

[1]- Institute of Cell Biophysics, RAS, Russia

[2]- University of Edinburgh, UK

## Running Head:

RKappa

## Corresponding Author:

Anatoly Sorokin

Institute of Cell Biophysics

Russian Academy of Sciences

142290, Puschino, Moscow Region

Russia

Email: **lptolik@gmail.com**

## Summary

RKappa is a framework for the development, simulation and analysis of rule-base models within the mature statistically empowered R environment. It is designed for model editing, parameter identification, simulation, sensitivity analysis and visualisation. The framework is optimised for high-performance computing platforms and facilitates analysis of large-scale systems biology models where knowledge of exact mechanisms is limited and parameter values are uncertain.

The RKappa software is an open source (GLP3 license) package for R, which is freely available online ( *https://github.com/lptolik/R4Kappa* ).

# 1. Introduction

The family of rule-based modelling techniques, including Kappa [1], BioNetGen [2], StochSim [3], etc. supports the formulation of mathematical models as collections of rules, which define allowable interactions and modifications [4]. Rule-based generalisation of interaction dynamics enables more effective scaling when compared to methods that consider each interaction independently and in full specificity/detail. This makes it a powerful approach when dealing with highly combinatorial and/or large-scale models of systems where precise knowledge of mechanisms and parameters is limited. If diffusion processes are not fast enough to justify a well-mixed assumption, it is possible to use the rule-based approach to account for spatial effects with tools such as SpatialKappa [5], SRSim [6], ChemCell [7], Smoldyn [8], and MCell [9].

The building and analysis of a dynamic model generally consists of several steps: model assembly, simulation, analysis and revision. As rule-based techniques are relatively new, only limited supporting infrastructure is currently available. For example, a Matlab-based library is available for BioNetGen, enabling parameter scanning, visualisation and analysis of simulation results [10], and a parameter fitting tool for BNGL-formatted models was developed only recently [11].

The RKappa package provides a first step towards a general pipeline that supports rule-based modelling techniques based around a Kappa simulator invoked within R. RKappa was developed around a central concept of parameter space exploration and global sensitivity analysis (GSA) [12]. This concept assumes that parameters vary widely in their effective influence within a model. GSA provides a method to rank the importance of parameters against their impact on behaviour of a deterministic or stochastic model. GSA and exploratory analysis of parameter space are the key steps in model development, as they facilitate an early identification of model predictions that do not align with available experimental data. It helps to reduce the number of parameters that need to be considered during a fitting procedure by excluding, or simplifying, non-influential parameters. It has also been shown that GSA of a parameterized model can provide important information for further experimental validation, development of a treatment strategy and identification of putative biomarkers [13].

The current implementation of RKappa supports sensitivity analysis in two modes, parallel and concurrent, which are selected depending on model structure and the purpose of a simulation.

In a more traditional parallel sensitivity experiment, parameter definition is separated from the rule, agent and observable definitions of a model allowing iterative substitution of parameter values over a range of interest during each simulation cycle. Thus, the number of individual models simulated equals the number of points sampled from parameter space. In concurrent sensitivity experiments, the rule, agent and observable definitions of a model are additionally divided into constant and variable subparts, where the constant part is independent on the parameters varied during sensitivity analysis. This feature was specifically designed for the situation where a group of similar agents could interact with a single target in a concurrent manner: for example, a transcription factor able to bind different parts of DNA with different affinities, or a phosphatase able to dephosphorylate several substrates with different efficiencies. Importantly, this feature allows us to assess parameter sensitivities when more than one element of a concurrent group is available and simultaneously takes part in an interaction. Unlike parallel

GSA, in the concurrent mode of operation we create a single supermodel by combining together different parameterizations of a model. This type of simulation is enabled by the compositionality property of rule-based models wherein a combination of two or more valid models becomes a valid model by itself. The capability to generate models and simulation jobs for concurrent GSA is a distinctive feature of RKappa. Examples of both models, parallel and concurrent are available from the RKappa website and described by Sorokin et al. [14].

Model analysis in RKappa starts with sampling the constrained available parameter space using the Sobol sampling procedure [15]. The actual number of samples depends on model size and complexity. For example, a model of 54 agents and 136 rules required 10000 parameter sets [16]. Criteria for selecting an appropriate number of simulation points are provided by Marino et al. [12] .

By default, RKappa performs GSA using a form of the partial rank correlation coefficient (PRCC) method [12] because of its moderate requirement for simulations with distinct parameter sets. However, eFAST, MPSA [17], and other algorithms for parameter set evaluation and sensitivity coefficient calculation, may be selected [18].

Given that many systems biology models have a high-dimensional parameter space, RKappa was designed to exploit high-performance computing (HPC) platforms, which facilitates computational analyses for large-scale models. The RKappa pipeline supports the distribution of computational tasks onto parallel computing resources, allowing simultaneous exploration of many parameter sets. We have simulated up to 50,000 parameter sets for a single model.

Here, we consider an RKappa-enabled analysis pipeline that consists of three main parts, running in a loop:

1) Initialisation of a numerical experiment,

2) Simulation based on a given parameter set, and

3) Analysis of the simulation results.

Parts 1 and 3 are computationally efficient and can be run on typical desktop workstations. Part 2 requires more computational resources and is best executed using HPC resources, with runs in a batch mode without manual intervention. A detailed step-by step description and use is presented in the Methods section.

## 2. Materials

RKappa is designed for and tested against rule-based models written in the Kappa language and simulated with KaSim versions from 2.5 to 4.0. Minor modifications would enable use of different simulation engines and languages such as BNGL and NFSim.  StochSim, for example may also be used in principle but facile application of RKappa workflows would require more extensive modification of the RKappa codebase.

The current release of the full RKappa pipeline was tested on Mac and Linux platforms. Scripts to support batch processing on HPC resources were tested with SunGrid Engine. Correct implementation of HPC invocation scripts can be done within the project definition setup steps, as shown later in Section 3.6, and more flexibility can be achieved with the use of the BatchJob R package [19].

RKappa generally assumes that a rule-based model is ready and valid in advance but it can be used to validate models ahead of runs on HPC resources, for which it has been optimized (see Section 3.5 below). For illustration purposes we use an example model of interactions among two trivalent agents and two divalent agents. A contact map for the model is shown on Figure 1. Trivalent agents A and B form hetero– and homopolymers of various lengths and structures, while divalent agents C and D decorate the polymer scaffolds. The size of a polymer is limited only by the number of available agents. This type of scenario cannot be represented by conventional models, such as ordinary differential equations, as the system has unlimited state space. It should be noted that the dynamics of the system are governed by certain parameter values  and initial conditions. Code for the model is shown in Figures 2, 3 and 4.

# 3. Methods

## 3.1. Installation of the package

The RKappa package is freely available from GitHub (*see* **Note 1**) and may be installed using the following R script:

```
install.packages ('devtools')

devtools::install_github('lptolik/R4Kappa')
```

```
library(rkappa)
```

## 3.2. Definition of parameter space

All parameters (reaction rates constants, concentrations, etc.) that influence the behaviour of a model should be divided into two categories: interesting, meaning that these parameters will be evaluated during further analysis; and 2) uninteresting, meaning that these parameters will be taken to have fixed values. Examples of uninteresting parameters may include the Avogadro constant, the cytoplasmic volume of a cell, or parameters found to have little influence on model predictions.

For interesting parameters (defined as variables in Kappa) a data frame needs to be generated that specifies the range of variation (name, min, max) (*see* **Note 2**):

```
prange <- data.frame(name = c('a', 'b', 'c'), min = rep(0,3),
max = c(10,20,50))
```

## 3.3. Model preparation

Rule –based model consists of definition of 1) agents (`Agent definition` on Figure 2), 2) rules (`Reaction rules` on Figure 2), which define possible interactions between the agents and modifications of agent states, 3) observables (`Observables` on Figure 2), which are the quantitative descriptions of system evolution, and 4) kinetic parameters  (Figure 3 and 4), which determine the timing of interactions [1,4]. To simulate the model, we create a predefined number of agents (`Initial condition definition` on Figure 2) and let system to evolve. The state of the system at a particular time point is defined by the number of agents in each their possible states and the complexes formed by interacting agents. We will use term "state" to refer to the state of an agent as a combination of its site variable values and interactions. To describe the state of the whole system we will use the term "reaction mixture". During simulation the application of rules will change the reaction mixture but its content at any time point can be saved as a snapshot for further analysis.

For sensitivity analysis the model should be restructured/split to three parts:

- A constant part. This part contains lines that do not depend on the variables defined in a model (Figure 2). Note that variable definitions are prefixed with "%var:" in a Kappa-formatted model.
- A template part. This part depends on variables but does not include their definitions (Figure 3).
- A parameters part. This part contains the definition of all parameters (Figure 4)

In this way for each simulation based on a distinct set of parameter values, a separate instance of the model will be created. For each instance, the constant part of the model will remain the same, whereas the template and parameter parts will vary (*see* **Note 3**). The names of the varying objects are appended with "_-" string, which will be substituted with the number/index of the corresponding parameter set. Names of parameter parts are updated in the same manner.

## 3.4. Generation of a simulation project.

A model needs to be embedded into a simulation project using the *prepareProject* command, which will specify the number of parameter sets to explore, the version of the simulator and the type of sensitivity analysis to perform. The following is an example of a *prepareProject* command.

```
abcdProj<-prepareProject(project='abcd',
numSets=50,
exec.path="~/kasim3/KaSim",
constantfiles=c('cABCD_const.ka'),
templatefiles=c("cABCD_templ.ka"),
paramfile=c("cABCD_param.ka"),
type='parallel')
```

In this command, the first argument/parameter *project* serves to define the name of the folder/ directory, where project executables will be placed. The second parameter, *numSets* specifies the number of parameter sets to be considered in simulations (*see* **Note 4**). The third parameter, *exec.path,* defines the path to the local installation of the simulator (KaSim,) on the platform being used. The parameters *constantfiles, templatefiles* and *paramfiles* define file names for the constant, template and parameter parts of model, respectively. The last parameter *type* takes one of three values: 'parrallel', 'concurrent' or 'both'. This parameter setting determines the type of GSA to be performed.

For simulation on an HPC platform the project requires three executable templates: *job.sh.templ, jobConc.sh.templ* and *run.sh.templ* (*see* **Note 5**).

By default these executable templates are predefined for the Sun grid engine under the Scientific Linux platform (*see* **Note 6**). If a different environment is required, the content of the project attributes must be modified appropriately. The *readFiles* function in the *rkappa* package may be used for this purpose. The files *Job.sh.templ* and *jobConc.sh.templ* define parallel execution and are specific for the cluster environment being used. The file *run.sh.templ* specifies the simulations with different parameter sets and should be modified only if particular parameters of KaSim have to be modified for correct model simulation. By default models are simulated for a certain amount of model time, as set by - t command line attribute (Figure 5). For some models it is more convenient to simulate for a certain number of events, which can be set using - e command line. In this case, the *run.sh.templ* file would be modified as shown in Figure 6.

## 3.5. Validation of newly generated project

To avoid unproductive, time-consuming and potentially costly error correction on compute clusters, models should be tested for syntactical and semantic errors locally (*see* **Note 7**). To support this RKappa can generate a few sample parameter sets and sequentially simulate them locally for a short time as a validation test. In addition to useful detection of basic syntactic errors in a model definition, this ensures that the project generation process does not result in a corrupt model structure and ensures compatibility with the chosen simulation engine. Other problems (such as state explosion or more subtle incorrect behavior of the model) cannot currently be detected by this initial model validation mode and these issues need to be identified by post-hoc analysis of simulation results. The command for this is as follows:

```
Out <- validate.kproject(abcdProj, dir='temp', nrep=2, nsets=1,
save=FALSE,
exe="~/kasim" )
```

where, `dir=tempdir()` defines an optional destination directory , and *nrep*=2 determines the number of repetitive evaluation calls to KaSim. A setting of 2 is recommended. The parameter setting *nsets*=1 defines the number of test parameter sets to evaluate. In cases where *nsets* > 1 validation is assumed to be performed in concurrent mode. The parameter setting `save=FALSE` is a logical statement that indicates whether to save results of the simulation (*see* **Note 8**). The parameter setting `exe=kproject$execPath` is an optional path, which may be different from the `execPath` of the main project. *Out* will contain simulation output generated by KaSim or other simulator and should be checked for error messages.

## 3.6. Preparation for HPC execution

Once models for each parameter set have been prepared, the scripts required to execute each simulation are generated. The pipeline is not tied rigidly to a specific simulation engine; instead it requires a user-defined template to invoke a selected simulation engine on the cluster. Default support for the KaSim3/KaSim4 engines are provided but the system has been tested using JSim and KaSim versions from 2.3 onwards.

```
write.kproject(abcdProj, projectdir = "ABCD")
```

This command will create the folder ABCD (by default the project name will be used), which willcontain all files for simulations with predefined parameter sets. This folder can be copied to the cluster being used and launched using *job.sh* or *jobconc.sh* scripts.

On the Sun grid engine the command for this is as follows:

```
qsub -t 1:50 job.sh,
```

which means that the first 50 parameter sets will be simulated in parallel.

During execution, results for each parameter set will be placed in folders named `pset1`, `pset2`, etc. folders. Because KaSim is a stochastic simulator, each parameter set will be simulated several times. The number of repeat simulations is specified by the `nRep` attribute of the project, which by default is 10. After execution has been completed, the content of the folders can be transferred back to a user's local machine for further analysis (*see* **Note 9**).

## 3.7. Loading and visualization of simulation results

The most simulation output is a time series reporting how a Kappa-defined observable varies over time (Figure 7).

The following shows how to load results from the local folder containing the results from the previous step using the `read.observables` command.

```
abcdObs<-read.observables(abcdProj,dir='abcd_res')
```

`The abcdObs` object contains four blocks of data:

- The `Project` block is a complete copy of the project. This block can be inspected to ensure that simulation parameter sets are consistent with user's analysis goals.

- The `resConc` block contains the results of concurrent simulations if any.

- The `resPar` block contains the results of parallel simulation. For each parameter set, it contains `nRep` timeseries for each trial.

- -The `aggPar block contains` aggregated (averaged) time series from `nRep` trials for each parameter set.

The `resPar` block can be visualised with following command from *ggplot2*:

```
qplot(time,AB,data=abcdObs$resPar[abcdObs$resPar$pset<=10,],face
ts = pset~.,geom = 'line',color=N,group=N).
```

Another type of simulation result is a snapshot, which is not commonly used but can be very informative. A snapshot reports the content of the reaction mixture at a chosen time point (*see* **Note 10**).

Snapshot data can be loaded with the command `read.snapshot.kproject` (*see* **Note 11**)*:*

```
snap<-read.snapshot.kproject(kproject = abcdProj,dir =
'abcd_res'),
```

`A snap` object contains 9 columns of data:

- "`T`" – time

- “`Ev`” – number of events
- “Num” – number of instances of a particular complex
- “`Kappa`” – the Kappa string representing a particular complex
- “`Brutto`” –the Brutto formula (from chemistry) of a particular complex, which indicates the number of instances of each agent type within the complex
- “`Weight`” – total number of agents within a particular complex
- “`Comp`” – total number of agent types within a particular complex
- “`Try`” and “`Set`” –the number of executions (`1:nRep`) and the index of a parameter set

For our example model, the information in the snapshot data frame is given in Table 1.

Table 1: First five lines of snapshot table

| Event | T | Num | Kappa | Brutto | Weight | Comp | Try | Set |
|---|---|---|---|---|---|---|---|---|
| 100 | 0.0105 | 4 | A(a,b,c!0),C(a!0,d) | A.1.C.1 | 2 | 2 | try1 | 10 |
| 100 | 0.0105 | 2 | A(a!0,b,c!1),A(a!0,b,c),C(a!1,d) | A.2.C.1 | 3 | 2 | try1 | 10 |
| 100 | 0.0105 | 3 | B(a,b!0,d),B(a,b!0,d) | B.2 | 2 | 1 | try1 | 10 |
| 100 | 0.0105 | 39 | B(a,b,d) | B.1 | 1 | 1 | try1 | 10 |
| 100 | 0.0105 | 88 | C(a,d) | C.1 | 1 | 1 | try1 | 10 |

Snapshots can be visualized using commands from the *igraph* package (Figure 8A). In a similar way we can plot any complex of interest, such as the largest one (Figure 8B).

## 3.8. Definition of metrics for sensitivity analysis

### 3.8.1 Observables metrics

To assess whether our model behaves properly we need to define metrics that quantify how far model behaviour is from a conceptual or data-driven ideal. During fitting, when experimental data are available, the usual choice for a cost function is dissimilarity between experimental data and simulation results. For time series obtained from observables with the proper time span, the cost function can be defined as follows:

```
cost<-function(data,expData){
    d<-sum((data-expData)^2)
}
```

In general the cost functions will be similar to those commonly used for fitting ODE models to experimental data [4] (*see* **Note 12**)

### 3.8.2 Snapshot metrics

Analysis of a snapshot can be very informative and it is where the power of rule based modeling with RKappa achieves its full potential. There are several steps in the analysis of snapshots, including analysis of mixture composition and analysis of complex structure formation/ disassembly.

For analysis of composition, we are interested in the number of complexes in a reaction mixture, their sizes and agent compositions. For example, if we are modelling polymerization, a key metric may be the size of a polymer. Such data are available from a *snapshot* object. Figure 9A shows distribution of complex sizes/weights obtained during simulation with the first 10 parameter sets of our *ABCD* model (*see* **Note 13**). We can see that on average the complexes are quite small.

If we are interested in the largest complex size (Figure 9B) we need to extract that data first. For this we will use the *plyr* package [20], as follows:

```
library(plyr)
maxWeight<-ddply(snap,
     .(Try,Set), summarize, Weight=max(Weight),
     Num=Num[which.max(Weight)], Comp=Comp[which.max(Weight)],
     Brutto=Brutto[which.max(Weight)],
     Kappa=Kappa[which.max(Weight)])
ggplot(maxWeight[maxWeight$Set<10,],
     aes(x=Set,y=Weight,group=Set)) +
     geom_boxplot()+
     scale_x_continuous(breaks=1:10)+
     ggtitle('Maximum Weigth polymers')
```

In a similar manner we can calculate the fraction of agents in the largest complex (*Comp*), or the number of complexes with the largest and smallest weights (*see* **Note 12**).

Another source of aggregated information is the Brutto string (*see* **Note 14**). For each complex, the Brutto string reports a count of each agent type. The following example code finds the complex that contains the larges amount of agent B (Figure 9C):

```
snap$nB<-as.integer(gsub('.*B\\.([0-9]+).*','\\1',snap$Brutto))
nB<-ddply(snap,.(Try,Set),
      summarize,
      totB=sum(nB*Num,na.rm=TRUE),
      ratB=max(nB*Num/totB,na.rm=TRUE),
      nB=max(nB,na.rm = TRUE))
maxB<-merge(snap,nB)
ggplot(maxB[maxB$Set<10,],
      aes(x=Set,y=ratB,group=Set))+
      geom_boxplot()+
      scale_x_continuous(breaks=1:10)+
      ggtitle('Portion of B agent in one complex')
```

### 3.8.3 Graph-based metrics

Frequently, the exact complex structure is of interest. To analyze complex structure we can use two functions: *makeIGraph* and *makeSiteGraph.*

```
sg<-makeSiteGraph("A(a, b, c!0),C(a!0, d)")
```

```
ig<-makeIGraph("A(a, b, c!0),C(a!0, d)")
```

The function *makeSiteGraph* produces a representation of a Kappa structure as a graph (Figure 10A) that accounts for all sites and bonds between sites. The function *makeIGraph* produces a simplified version of the complete site graph (Figure 10B), in which all bonds between sites of two agents are collapsed into a single edge (*see* **Note 15**).

Now if we can formulate the ideal behavior in terms of graph properties, we can create a graph-based cost function. For example, we can count the number of particular subcomplexes (Figure 11A). A complex containing a subgraph isomorphic to a pattern is shown in Figure 11B.

```
ab2<-makeIGraph("A(a!1,b!2),B(a!2,b!3),A(a!1,b!4),B(a!4,b!3)")
snap$AB2<-sapply(snap$Kappa,function(.x){
     g<-makeIGraph(.x);
     d<-lapply(V(ab2)$name,function(.x) which(V(g)$name==.x));
     return(count_subgraph_isomorphisms(ab2,g,
          method = 'lad',domain=d))
     })
```

It is important for subgraph isomorphism to take into account names of agents, so patterns are reproduced exactly. The IGraph function `count_subgraph_isomorphisms` uses the `domain` parameter for this purpose. It should contain a common list of vertices in the target graph, which represent the same agent. It should be noticed that because of the symmetry of the pattern *ab2* its count is always even (*see* **Note 16**).

## 3.9. PRCC and ACE-PRCC sensitivity analysis

### 3.9.1 Time-invariant metrics

Once a set of metrics is prepared and their values are calculated we can assess how much influence each parameter has on the metric's value. We can calculate the partial rank correlation coefficient (PRCC) as proposed in [12] by averaging selected metrics over simulation repeats (tries) for each parameter set of interest. With a previously calculated object nB this can be done as follows:

```
avgnB<-ddply(nB,.(Set),summarize,
     ratB=mean(ratB),
     nB=mean(nB))
```

Now we can calculate sensitivity coefficients as follows:

```
ps<-parallelsensitivity(abcdProj$paramSet,
     avgnB, outName='B',
     nboot=0)
```

Here, the first argument (`abcdProj$paramSet`) is the data frame with parameter values, and the second argument (`avgnB`) is the metrics data frame. The metrics name template is in `outName`. Results of the calculation contain three fields: `N` — the number of parameter sets

used for calculation, `p` — the number of excluded variables, and `prcc` — the data frame with the sensitivity analysis results. The object `prcc` contains three columns for each metrics: `sc`.* — the sensitivity coefficient, `T.*` - statistics used to calculate a significance value, and `pval` — *p*-value for significance (*see* **Note 17**). See Table 2.

Table 2: The results of PRCC calculations

| Parameter | sc.ratB | T.ratB | Pval.ratB | Sc.nB | T.nB | Pval.nB |
|---|---|---|---|---|---|---|
| MOD | 0.2629558 | 1.766309 | 0.0846117 | 0.7873578 | 8.2768737 | 2.3055e-1 |
| BRK | -0.4680712 | -3.4327037 | 0.0013549 | -0.807024 | -8.856815 | 3.6941e-1 |
| aa | -0.0010124 | -0.0065614 | 0.9947958 | 0.3932 | 2.7714608 | 8.2807e-0 |
| bb | 0.0376252 | 0.2440123 | 0.8084093 | 0.4090996 | 2.9055324 | 5.8272e-0 |
| ab | 0.0227919 | 0.1477470 | 0.8832494 | 0.4950745 | 3.6927520 | 6.3426e-0 |
| ac | -0.2070986 | -1.3718951 | 0.1773800 | -0.1463535 | -0.9588037 | 3.4314e-0 |
| bd | 0.0990867 | 0.6453312 | 0.5222225 | 0.0694929 | 0.4514571 | 6.5398e-0 |

Analysis of `pval` shows that for ratB there is only one parameter (BRK) significant at the level of 1%, whereas for nB there are five of such parameters. It should be noted that the most sensitive parameter, BRK, has a negative sensitivity coefficient, which means it is negatively correlated with the metrics.

The key limitation of PRCC method is that it requires a monotonic dependency between parameters and metrics. When dependency is non-monotonic, the PRCC value quickly changes towards zero.  In the case where a sensitivity coefficient is small, we can transform both parameters and metrics to obtain maximum linearity using the ACE method [21]:

```
library(acepack)

acenB<-ace(abcdObs$project$paramSets,avgnB$ratB)
```

```
qplot(acenB$y,acenB$ty,geom = 'line')

qplot(acenB$x[2,],acenB$tx[,2],geom = 'line')

qplot(acenB$x[3,],acenB$tx[,3],geom = 'line')
```

Figure 12 illustrates that the ratB metric itself is quite monotonic (A), while BRK is weakly non-monotonic (B), and aa is strongly non-monotonic (C).

Given these findings we recalculate PRCC with ACE-transformed parameters as follows:

```
ace.ps<-parallel.sensitivity(
     as.data.frame(acenB$tx),
     data.frame(ratB=acenB$ty),
     outName = 'B')
```

As can be seen in Table 3, sensitivity values are now much more significant. The non-monotonic dependency of *ratB* from *aa* parameter can explain extremely low PRCC value for *aa* in the Table 2, and its higher although non-significant sensitivity constant in the Table 3. The sensitivity coefficients for parameter *ab* become significant after transformation, which makes much more sense from the structure of the model.

Table 3: Result of ace-PRCC analysis

| Parameter | sc.ratB | T.ratB | pval.ratB |
|---|---|---|---|
| MOD | 0.4114281 | 2.925428 | 5.527167e-03 |
| BRK | 0.6567341 | 5.643811 | 1.290323e-06 |
| aa | 0.1863914 | 1.229501 | 2.257276e-01 |
| bb | 0.1964160 | 1.298209 | 2.012993e-01 |
| ab | 0.3557936 | 2.467252 | 1.777529e-02 |
| ac | 0.4878442 | 3.621812 | 7.819837e-04 |
| bd | 0.3253610 | 2.229910 | 3.115498e-02 |

### 3.9.2 Time-varying metrics

Sensitivity coefficients can be calculated for metrics that change with time. By plotting the value of sensitivity coefficient as a function of time we can identify the most and least sensitive time points forthe metrics of interest. RKappa analysis of time-varying PRCC can be performed with the following command:

```
res<-ddply(abcdObs$resPar,
     .(time,pset),summarize,
     avgAB2=mean(AB2),
     avgAA=mean(AA),
     avgBB=mean(BB),
     avgAB=mean(AB))
tps<-timed.parallel.sensitivity(abcdObs$project$paramSet,
     res, outName='avg', nboot=0)
```

To make a more informative plot we add the value of 1% significance level for sensitivity coefficients:

```
threshold<-prccConfidenceInterval(c(1e-3,5e-3,1e-2,5e-2,1e-1),
     attributes(tps)$N,
     attributes(tps)$p)
```

Now we plot the obtained sensitivity coefficients (Figure 13):

```
qplot(time,sc.avgBB,data = tps,
     group=param, geom='line', color=param)+
     geom_hline(yintercept = threshold$lower[3])+
     geom_hline(yintercept = threshold$upper[3])
```

## 3.10. Updating sampling of parameter space and sensitivity refinement

In many cases, a low significance of a sensitivity value means that an insufficient number of parameter sets was analyzed (*see* **Note 18**). By using the Sobol low-discrepancy sequence to generate parameter sets it is possible to add more parameter sets without replacing existening ones and tereby minimize the number of necessary calculations (*see* **Note 19**):

```
abcdProj<-addSets(abcdProj,nStart =
51,nSets=100,seed=abcdProj$seed)
```

Here `nStart` is the index of the first parameter set to be replaced, `nSets` is the number of new sets to be generated, and seed is the pseudo random generator seed.

Now one can write the project into the same folder and simulate the model for newly generated parameter sets.

# 4. Notes

**1.** For historical reasons, the RKappa project on GitHub is called *`R4Kappa`* and the library in R is called *`rkappa`*

**2.** There is another way to define parameter ranges: by providing multipliers for parameters in the model. In this case names of the variables will be taken from the parameter part (see the next step) and their ranges will be modified by multiplication by the specified *min* and *max* values.

In the parameter data frame, the number of columns should match the number of free-to-vary parameters, while the number of rows should be equal, or greater than, `nSets`. If `nrow` is less than `nSets` an error message will be generated upon the attempt to create a project, and the project folder will not be created.

**3.** This model subdivision is most important when concurrent sensitivity is selected, as not only the variables names but also the agent names will be updated.

**4.** The number of points in parameter space that will be chosen for simulation should be less than, or equal to, the number of rows in the parameter data frame. If *`nSets > paramSets`*, the attempt to export the project to the executable folder will result in an error. In this case additional rows can be created with the command "`addSets`".

**5.** There are several predefined templates: `run.sh.templ`,`job.sh.templ`, and `jobConc.sh`. These templates are described in turn below.

`run.sh.templ`

This script defines steps to evaluate one parameter set. The script repeats simulation of the specified model several times and stores results in a predefined set of folders. It is this script that defines the simulation engine and it has to be modified for use with different versions of KaSim or other simulation engine , such as NFsim for example.

`job.sh.templ`

This is the whole job execution script. The template provided with the package is designed for execution with the Sun Grid Engine batch-queuing system and was tested on the ECDF Cluster at Edinburgh University. Execution of the jobs on other systems such as SLURM, HTCondor or in AmazonEC2 will require some modification of the template.

`jobConc.sh`

This script is the concurrent sensitivity job execution script. The template provided with the package was designed for execution with the Sun Grid Engine batch-queuing system and was tested on the ECDF Cluster at Edinburgh University. Execution of jobs on other system will require some modification of the script; see comments above..

**6.** To simulate the content of the project locally you should define the path to local executables as a separate parameter.

**7.** The most common reason for errors are typos and incomplete update of the template part during the model split. For example, an error will occur if a parameter name has been changed in the parameter definition but the name was not updated accordingly in the rule specification part. It is important that the version of the simulator available on the HPC platform be as close as possible to the version on the local machine, as error fixing on the HPC platform could be time-consuming and a waste of HPC resources.

**8.** It could be useful to switch save to TRUE if validation fails to ensure that export of the project is complete and all scripts are ready for simulation.

**9.** Analysis of snapshots may be time and memory consuming so further rounds on the HPC platform may be required.

**10.** The frequency of snapshot taking can be specified within a Kappa-formatted model with perturbation semantics as follows:

```
%mod: repeat ([E+] > 1) && ([E+] [mod] 100)=0 do $SNAPSHOT
"cABCD" until [false] - for every 100 events
```

These commands indicate the snapshot will be generated for every 100 events.

```
%mod: repeat ([T] > 0) && ([T] [mod] 1.0)=0 do $SNAPSHOT "cABCD"
until [false]-
```

These commands indicate that snapshot will be generated for each second of model time.

**11.** If snapshots are taken too often, the size of the results can grow substantially (a cumulative file size over 10Gb would not be unusual). The function `read.snapshot.kproject` has two optional parameters `getAll` and `sfile`, which may be used to load only the required parts of snapshots. By default `getAll` is TRUE, which means that the function will try to load all snapshots generated during a simulation. *S*etting `getAll` to FALSE will inform the system to load only the last snapshot, which will correspond to the highest event count suffix. To load only the last snapshot a user may call the function in the following way:

```
snap<-read.snapshot.kproject(kproject = abcdProj,dir =
'abcd_res', getAll=FALSE)
```

Another approach to reduce the number of generated and/or analysed snapshots is to generate snapshots with a particular naming structure and read only those files via the *sfile* parameter. For example, KaSim provides a command to store a snapshot of a reaction mixture under certain conditions:

```
%def: "dumpIfDeadlocked" "yes"
```

This command will create the file "`deadlock.ka`" when there is no rules can be applied to the reaction mixture. To load only the deadlocked files produced by a simulation you can execute the following command:

```
snap<-read.snapshot.kproject(kproject = abcdProj,dir =
'abcd_res',sfile='deadlock.ka')
```

**12.** RKappa does not provide predefined cost functions within a project. We note that KaSim is stochastic simulator; thus, the stochastic nature of simulation result shave to be addressed properly. It is for the user to decide how to account for the differences between individual simulation runs. However, R is an excellent environment for this type of situation because of the number of available packages dedicated to dealing with stochastic data.

**13.** To prepare the plot, execute the following commands:

```
library(ggplot2)
ggplot(snap[snap$Set<10,],
    aes(x=Set,y=Weight,group=Set))+
    geom_boxplot()+
    scale_x_continuous(breaks=1:10)+
    ggtitle('Polymer weigth distribution')
```

**14.** It should be noted that we can assess not only the value of characteristics like weight of a complex but also various statistics of it. For example, in the Figure 8C, the fifth parameter set has the largest range and interquartile distance, while the first and the seventh parameter sets have outliers.

**15.** Both `makeIGraph` and `makeSiteGraph` can accept lists of multiple Kappa strings. In this case the functions attempt to reproduce the reaction mixture, with the aid of the optional parameter *num* to store the number of a particular complex in a reaction mixture.

**16.** If snapshots are taken at different time points we can analyse how the graph properties of a reaction mixture changes with time. Analysis of such large series of graphs can be extremely time consuming and may require use HPC resources.

**17.** The most unusual parameter is `outName`: it specifies a regular expression, which is used to select metrices for analysis. Details of regular expression use are not important here. Using a name substring is typically adequate. For example in our case `outName='B'` will match both *ratB* and *nB*.

**18.** In the case where metric dependency is described by non-monotonic function, the PRCC coefficient could be insignificant. In this case an increase of parameter set number could lead to a decrease in the sensitivity coefficient value.

**19.** If you add new parameters it is important to use the same seed, so the all parameters will be from the same sequence of pseudorandom numbers. If you want to completely replace all parameter sets use *`nStart=1`*, so that new parameters will be generated. However, but in this case, you should write the project to a new directory; the default `run.sh` script does not execute new KaSim runs if results of a previous run already exist.

## Figure Captions

**Figure 1.** Contact map for the model. There are four agents in the model, two of which (A and B) are trivalent, and two (C, D) are divalent. Agent A has three sites a, b and c. Site A.a is responsible for dimerisation of A agents, site A.b can form a bond with site B.a of agent B. Site A.c is responsible for binding with agent C via bond with site C.a. Similarly, for agent B site B.b is responsible for dimerisation and site B.d for binding with agent D via site D.b. Sites C.d and D.d of agents C and D respectively do not participate in bond formation.

**Figure 2.** Constant part of the model. The constant part of the model will not be affected by changes during sensitivity analysis. It consists of four parts: agent definitions, rule definitions, the list of observables, a snapshot definition and initial conditions. None of these parts explicitly reference free parameters, which have names ending with "_-".

**Figure 3.** Template parameter definition part of the model. The template part of the model lists all variables, rules and observables that explicitly depend on free parameters (with names ending with "_-"). Note that we made all kinetic constants the variables that are calculated from free parameters, which enables moving the rule definitions into the constant part of the model.

**Figure 4.** Free parameter definition (parameters) part of the model. RKappa substitutes parameter values in this part of the model with values from a particular parameter set. Values

provided in this part could be useful for testing purposes. They are also used if parameter ranges are specified in terms of relative values (*see* **Note 2**).

**Figure 5.** Parameter set evaluation script template to run a model for a given period of time. After substitution of %nSet% and %simTime% with values from the project copy of that script will be created for each parameter set (%setIndx% to be replaced by number of the set).

**Figure 6.** Parameter set evaluation script to run a model for a given number of reaction events. The same as Figure 5, except %simTime% will define the number of events rather than the time.

**Figure 7.** Time series of AB observable for the first 10 parameter sets. Each panel represents `nrep` evaluations of the parameter sets. It can be seen that variability in AB behaviour is higher in panels 8 and 4 and is much lower in panels 5 and 7.

**Figure 8.** Graph representation of the snapshot data. A. Full reaction mixture. All complexes found in the snapshot will be copied in accordance with the number of complex instances, and displayed (with automatic layout). At the beginning of a simulation, the reaction mixture consists of individual agents only. We can create a movie of the evolution of the reaction mixture by combining consecutive snapshot graphs. B. The largest complex. It is often interesting to analyse the structure of only one complex. In this panel we draw the largest one.

**Figure 9.** Box plot of snapshot analysis data. The box plot provides the same data as a histogram but in a more concise way, so it is useful for comparison of different datasets. The top and bottom line of a box represents the upper and lower quartile of a distribution, respectively. The horizontal line is located at the median value and the whiskers show outlier boundaries. Dots represent outliers. A. Weight - size of the polymer. B. MaxWeight - size of the largest complex, C. ratB - proportion of B agents in the complex

**Figure 10.** Two representations of the Kappa string "A (a, b, c!0),C(a!0, d)". A. Representation created by using the `makeSiteGraph` command. Thick lines connect the central agent node with smaller site nodes. The thin line shows the binding between sites. Agent A is connected to agent C via A.c to C.a binding. B. Representation created by using the `makeIGraph` command. In this representation we only preserve complex structure at the agent level. All sites are hidden and even if there are several parallel bonds between two agents they will be shown with just one line.

**Figure 11.** Search pattern (A) and matched structure (B)

**Figure 12.** ACE transformation for analysis of ratB dependency from BRK and aa. A. ratB after transformation; B. BRK after transformation. A monotonic non-linear dependency is observe. C. aa — strong non-monotonicity is observed.

**Figure 13.** Time-varying sensitivity for BB observable. The horizontal lines define a low significance interval. We can see that three parameters become significant at the end of simulation; none are significant at the beginning. This type of plots makes it easier to plan data sampling time points for both real and computational experiments.

## Table Captions

**Table 1.** First five lines of the snapshot table

**Table 2.** The results of PRCC sensitivity analysis for metrics ratB and nB

**Table 3.** The results of ace-PRCC sensitivity analysis for metric ratB

## Tables

**Table 1**

| Event | T | Num | Kappa | Brutto | Weight | Comp | Try | Set |
|---|---|---|---|---|---|---|---|---|
| 100 | 0.0105 | 4 | A(a,b,c!0),C(a!0,d) | A.1.C.1 | 2 | 2 | try1 | 10 |
| 100 | 0.0105 | 2 | A(a!0,b,c!1),A(a!0,b,c),C(a!1,d) | A.2.C.1 | 3 | 2 | try1 | 10 |
| 100 | 0.0105 | 3 | B(a,b!0,d),B(a,b!0,d) | B.2 | 2 | 1 | try1 | 10 |
| 100 | 0.0105 | 39 | B(a,b,d) | B.1 | 1 | 1 | try1 | 10 |
| 100 | 0.0105 | 88 | C(a,d) | C.1 | 1 | 1 | try1 | 10 |

**Table 2**

| Parameter | sc.ratB | T.ratB | Pval.ratB | Sc.nB | T.nB | Pval.nB |
|---|---|---|---|---|---|---|
| MOD | 0.2629558 | 1.766309 | 0.0846117 | 0.7873578 | 8.2768737 | 2.3055e-10 |
| BRK | -0.4680712 | -3.4327037 | 0.0013549 | -0.807024 | -8.856815 | 3.6941e-11 |
| aa | -0.0010124 | -0.0065614 | 0.9947958 | 0.3932 | 2.7714608 | 8.2807e-03 |
| bb | 0.0376252 | 0.2440123 | 0.8084093 | 0.4090996 | 2.9055324 | 5.8272e-03 |
| ab | 0.0227919 | 0.1477470 | 0.8832494 | 0.4950745 | 3.6927520 | 6.3426e-04 |
| ac | -0.2070986 | -1.3718951 | 0.1773800 | -0.1463535 | -0.9588037 | 3.4314e-01 |
| bd | 0.0990867 | 0.6453312 | 0.5222225 | 0.0694929 | 0.4514571 | 6.5398e-01 |

**Table 3**

| Parameter | sc.ratB | T.ratB | pval.ratB |
|---|---|---|---|
| MOD | 0.4114281 | 2.925428 | 5.527167e-03 |
| BRK | 0.6567341 | 5.643811 | 1.290323e-06 |
| aa | 0.1863914 | 1.229501 | 2.257276e-01 |
| bb | 0.1964160 | 1.298209 | 2.012993e-01 |
| ab | 0.3557936 | 2.467252 | 1.777529e-02 |
| ac | 0.4878442 | 3.621812 | 7.819837e-04 |
| bd | 0.3253610 | 2.229910 | 3.115498e-02 |

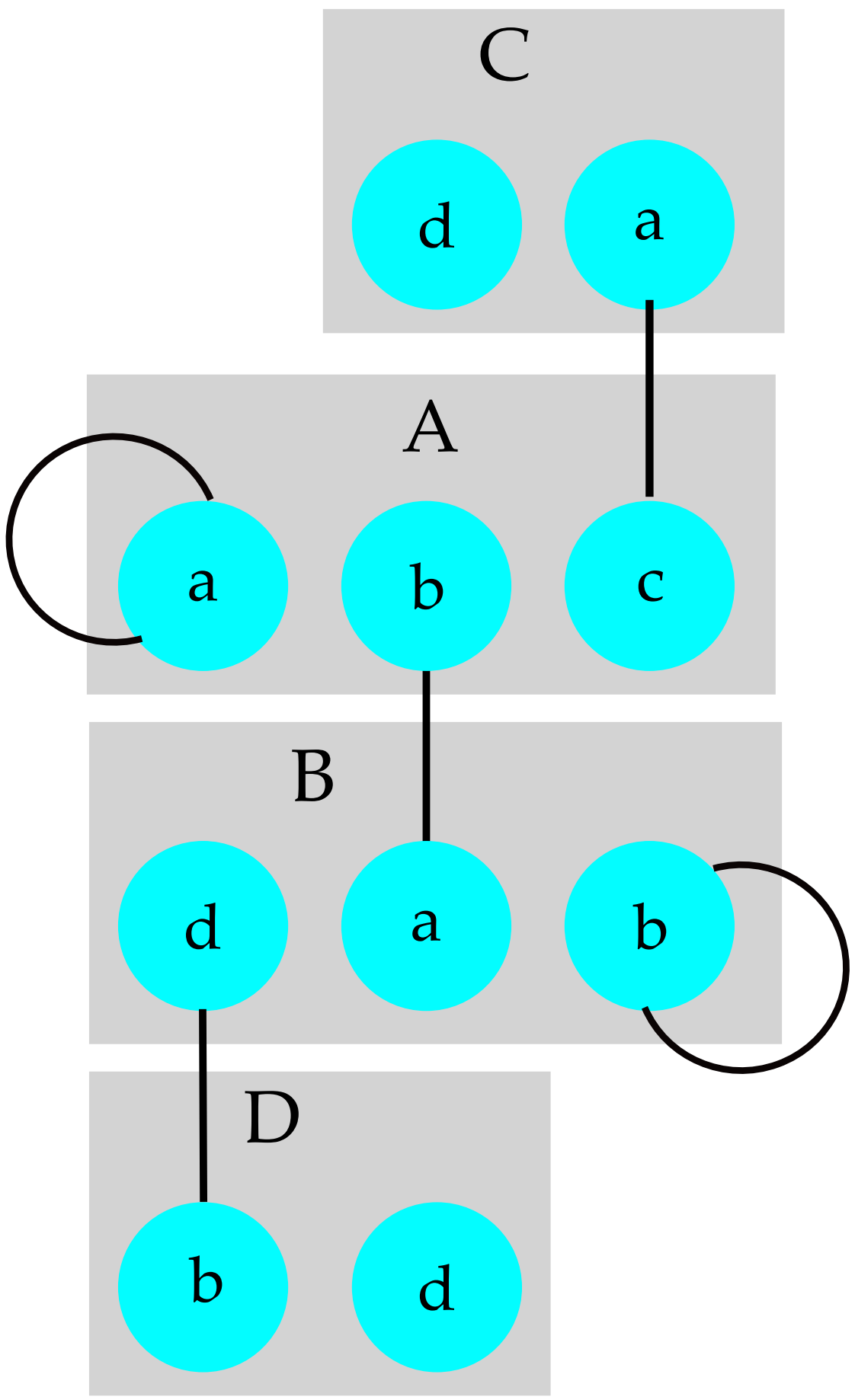

Figure 1.

```
#Agent definition
%agent: A(a,b,c)
%agent: B(a,b,d)
%agent: C(a,d)
%agent: D(b,c)

# Reaction rules
'assocAB'  A(b), B(a) -> A(b!0), B(a!0) @ 'BNDab'
'dissocAB' A(b!0), B(a!0) -> A(b), B(a) @ 'BRK'
'assocAA'  A(a), A(a) -> A(a!0), A(a!0) @ 'BNDaa'
'dissocAA' A(a!0), A(a!0) -> A(a), A(a) @ 'BRK'
'assocBB'  B(b), B(b) -> B(b!0), B(b!0) @ 'BNDbb'
'dissocBB' B(b!0), B(b!0) -> B(b), B(b) @ 'BRK'

'assocAC'  A(c), C(a) -> A(c!0), C(a!0) @ 'BNDac'
'dissocAC' A(c!0), C(a!0) -> A(c), C(a) @ 'BRK'

'assocBD'  B(d), D(b) -> B(d!0), D(b!0) @ 'BNDbd'
'dissocBD' B(d!0), D(b!0) -> B(d), D(b) @ 'BRK'

# Observables
%obs: 'Ev' [E+]
%obs: 'AB2' A(a!1,b!2),B(a!2,b!3),A(a!1,b!4),B(a!
4,b!3)

# Snapshot definition
%mod: repeat ([E+] > 1) && ([E+] [mod] 100)=0 do
$SNAPSHOT "cABCD" until [false]

# Initial conditions definition
%init: 'nA' A(a,b,c)
%init: 'nB' B(a,b,d)
%init: 'nC' C(a,d)
%init: 'nD' D(b,c)
```

Figure 2

```
# Template parameters definition
%var: 'vol' 0.1
%var: 'nA'  1000 * 'vol'
%var: 'nB'  1000 * 'vol'
%var: 'nC'  1000 * 'vol'
%var: 'nD'  1000 * 'vol'
%var: 'BND' 'MOD_-' / 'vol'
%var: 'BNDaa' 'BND' * 'aa_-'
%var: 'BNDbb' 'BND' * 'bb_-'
%var: 'BNDab' 'BND' * 'ab_-'
%var: 'BNDac' 'BND' * 'ac_-'
%var: 'BNDbd' 'BND' * 'bd_-'
# Observables
%obs: 'AA' A(a!_)
%obs: 'BB' B(b!_)
%obs: 'AB' B(a!_)
```

Figure 3

```
# Reaction kinetic parameters
%var: 'MOD_-' 0.0005
%var: 'BRK_-' 0.1
%var: 'aa_-' 1
%var: 'bb_-' 1
%var: 'ab_-' 1
%var: 'ac_-' 1
%var: 'bd_-' 1
```

Figure 4

```
#!/bin/bash
numEv=%nRep%
time=%simTime%
out=100
if [ "$1" != "" ]; then
numEv=$1
echo "number of events to simulate=$numEv"
fi
if [ "$2" != "" ]; then
time=$2
echo "number of seconsd to simulate=$time"
fi

i=1
echo $i
[ ! -d "./pset%setIndx%/try$i" ] && mkdir -p "./pset%setIndx%/try$i"
cd "./pset%setIndx%/try$i"
if [ ! -f ./data.out ]
then
    echo "KASIM_EXE  try$i -t $time -p $out -d "./" > out.txt"
    $KASIM_EXE %inputs% -t $time -p $out -d "./" > out.txt
fi
cd ../
while [ $i -lt $numEv ]
do
i=$[$i+1]
mkdir -p "./try$i"
cd "./try$i"
if [ ! -f ./data.out ]
then
    echo "$KASIM_EXE  try$i -t $time -p $out -d "./" > out.txt"
    $KASIM_EXE %inputs% -t $time -p $out -d "./" > out.txt
fi
cd ..
done
cd ..
```

Figure 5

```
#!/bin/bash
numEv=%nRep%
time=%simTime%
out=100
if [ "$1" != "" ]; then
numEv=$1
echo "number of events to simulate=$numEv"
fi
if [ "$2" != "" ]; then
time=$2
echo "number of seconsd to simulate=$time"
fi

i=1
echo $i
[ ! -d "./pset%setIndx%/try$i" ] && mkdir -p "./pset%setIndx%/try$i"
cd "./pset%setIndx%/try$i"
if [ ! -f ./data.out ]
then
    echo "KASIM_EXE  try$i -e $time -p $out -d "./" > out.txt"
    $KASIM_EXE %inputs% -e $time -p $out -d "./" > out.txt
fi
cd ../
while [ $i -lt $numEv ]
do
i=$[$i+1]
mkdir -p "./try$i"
cd "./try$i"
if [ ! -f ./data.out ]
then
    echo "$KASIM_EXE  try$i -e $time -p $out -d "./" > out.txt"
    $KASIM_EXE %inputs% -e $time -p $out -d "./" > out.txt
fi
cd ..
done
cd ..
```

Figure 6

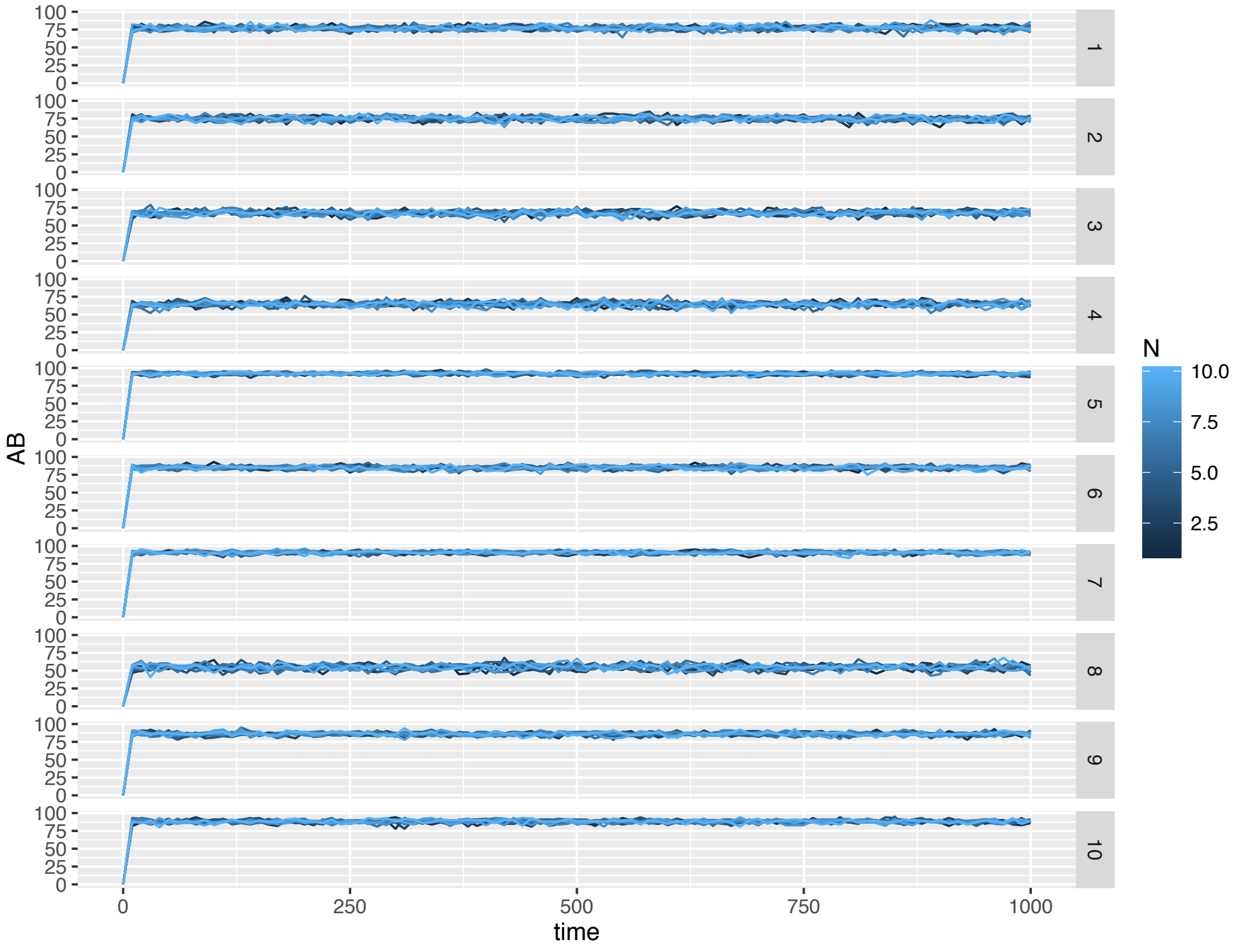

Figure 7

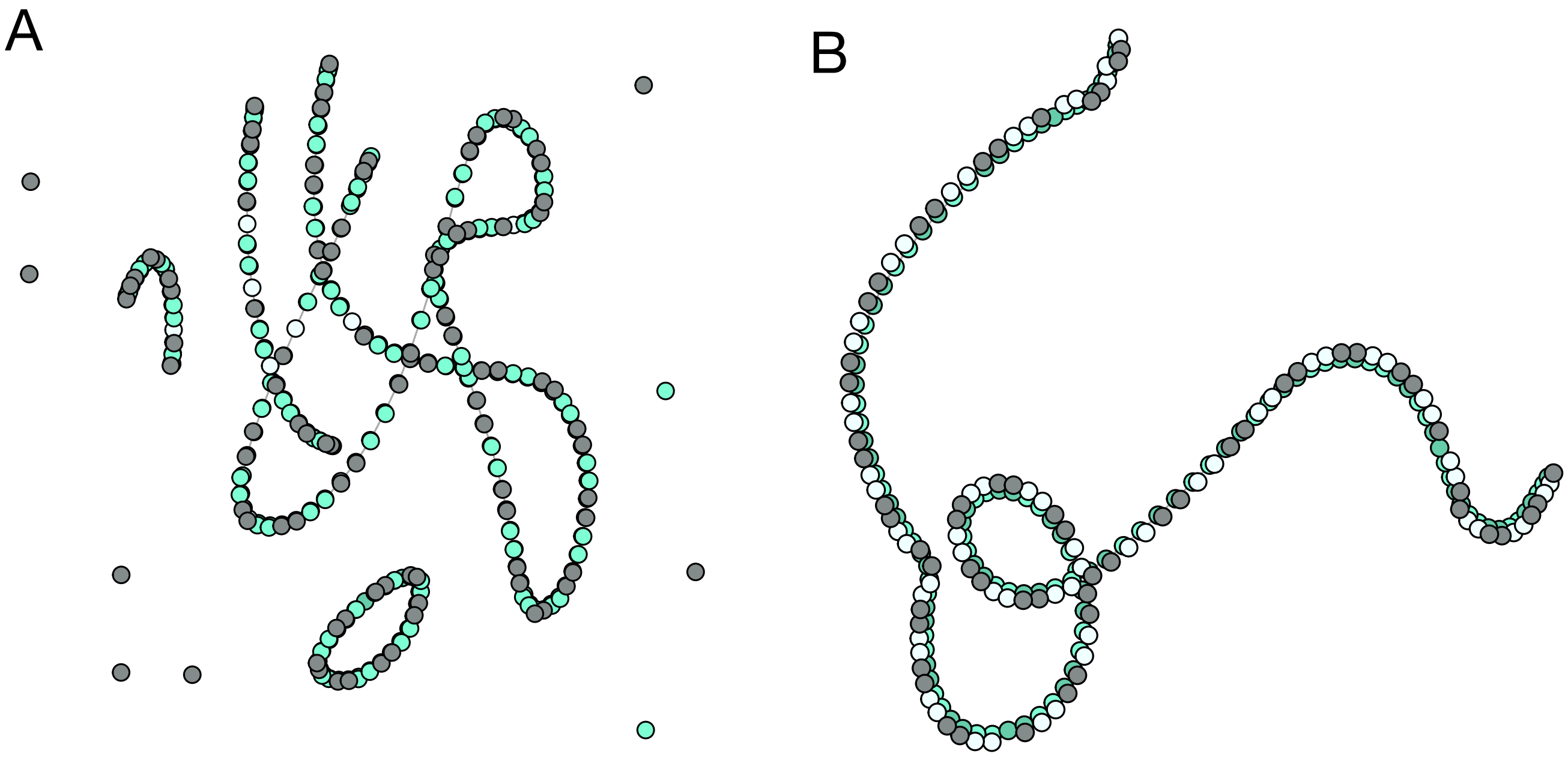

Figure 8

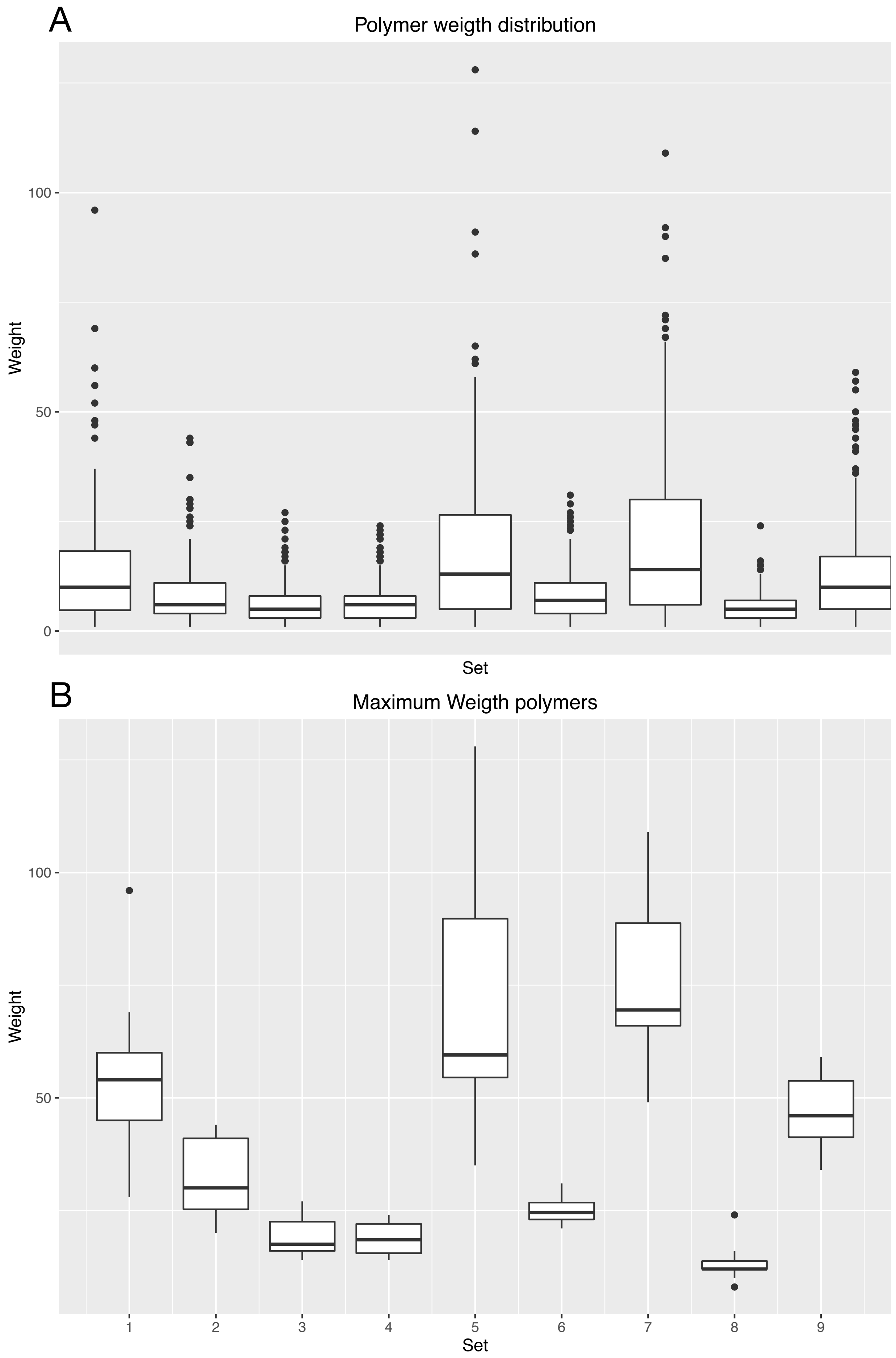

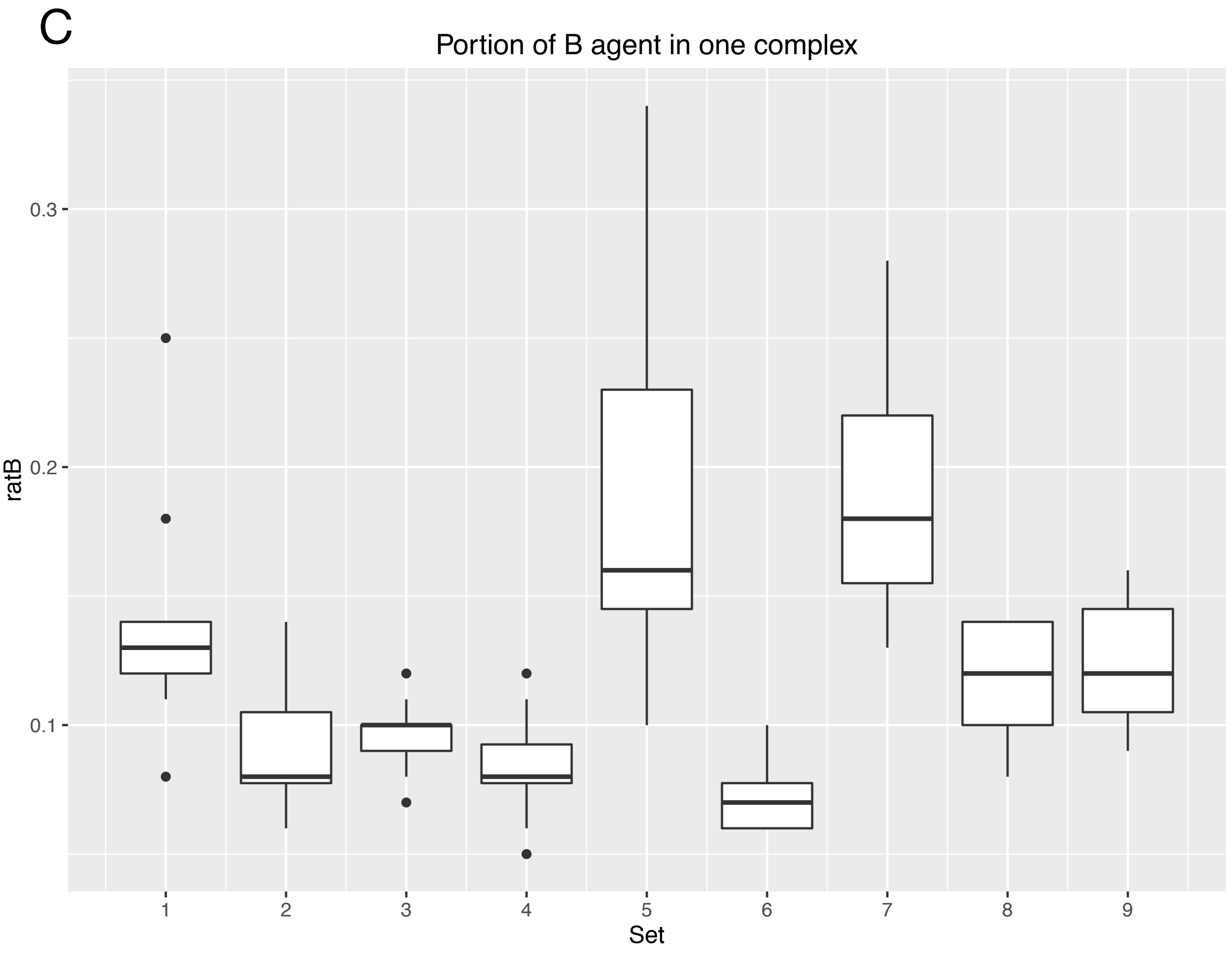

Figure 9

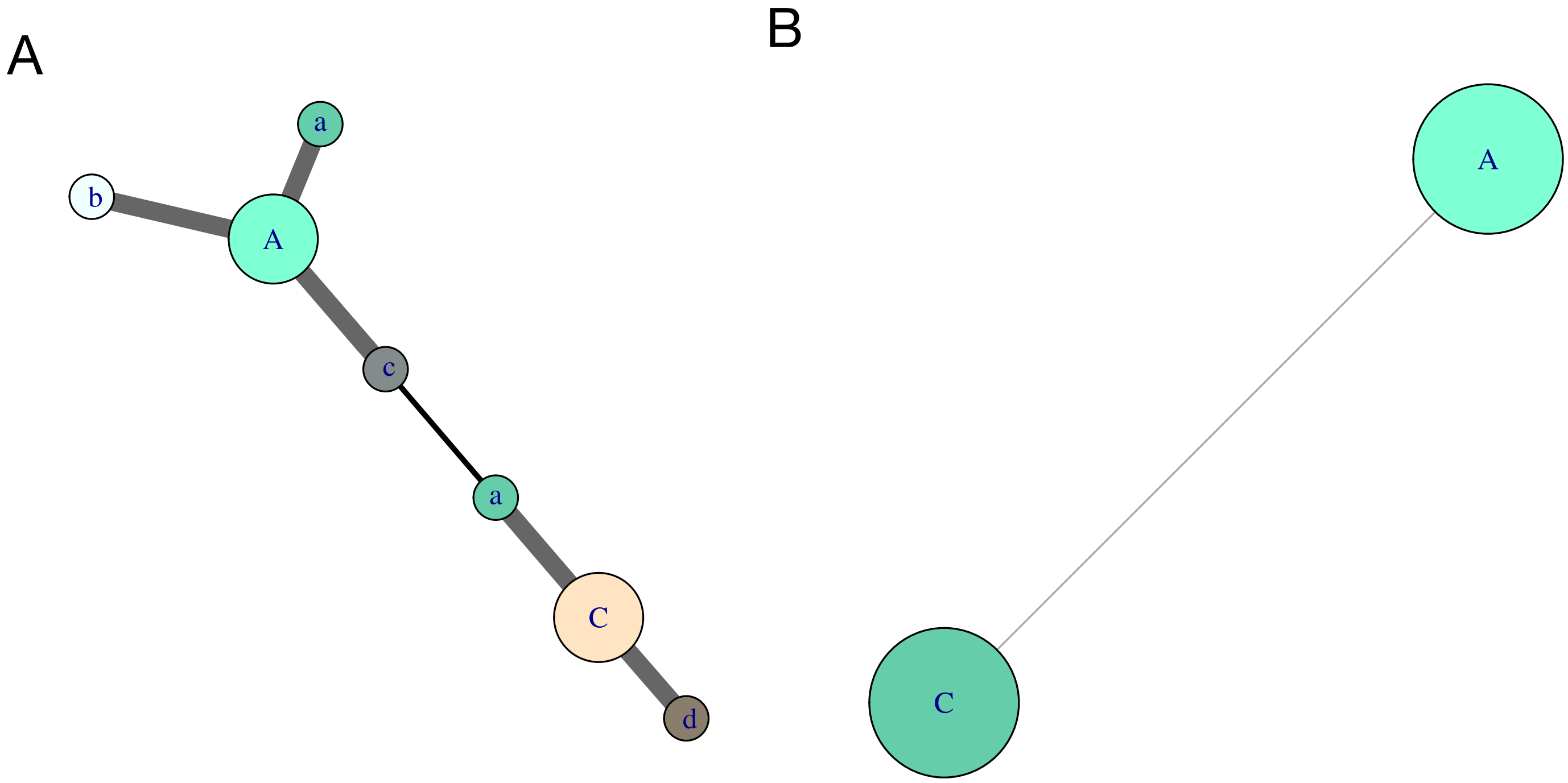

Figure 10

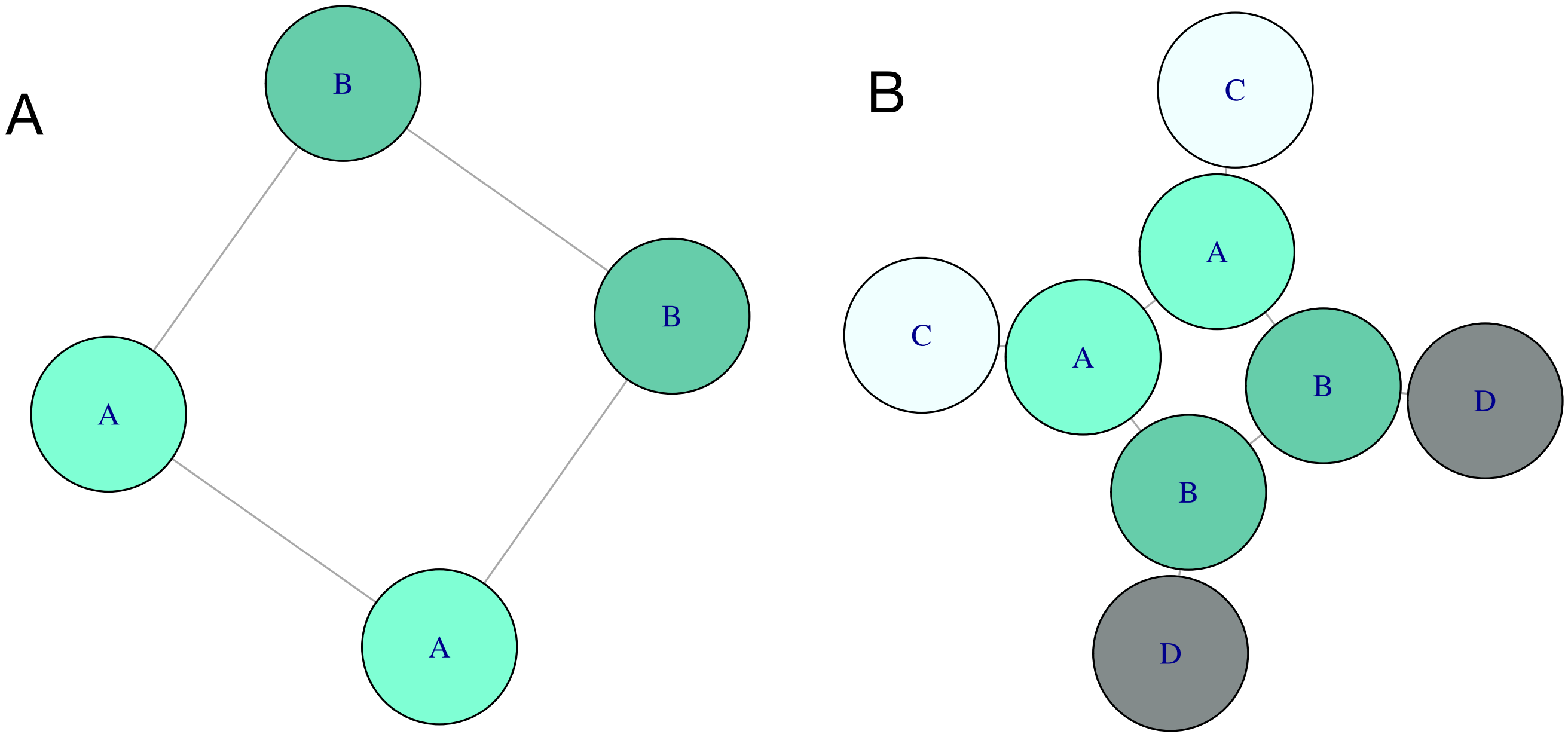

Figure 11

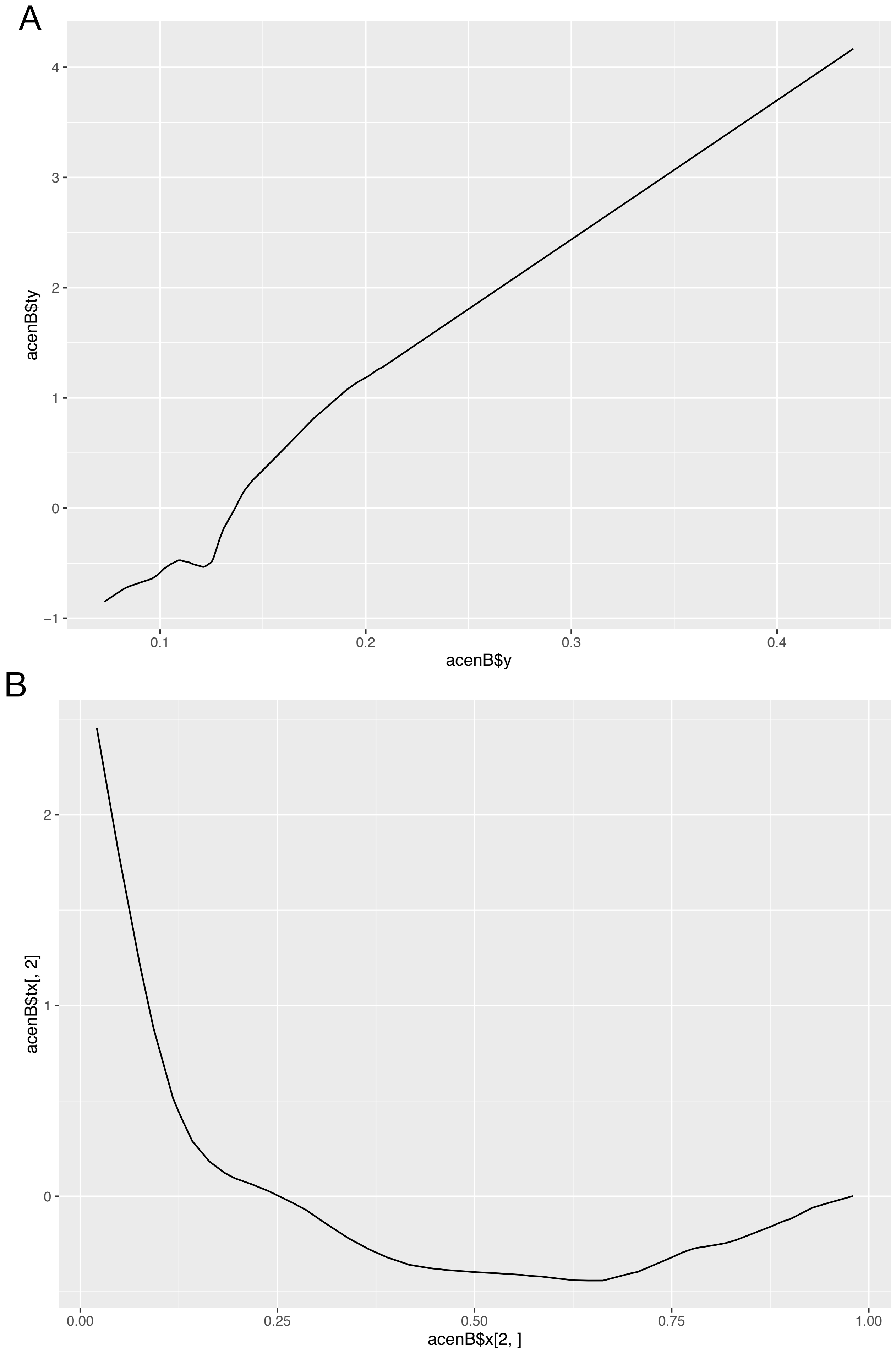

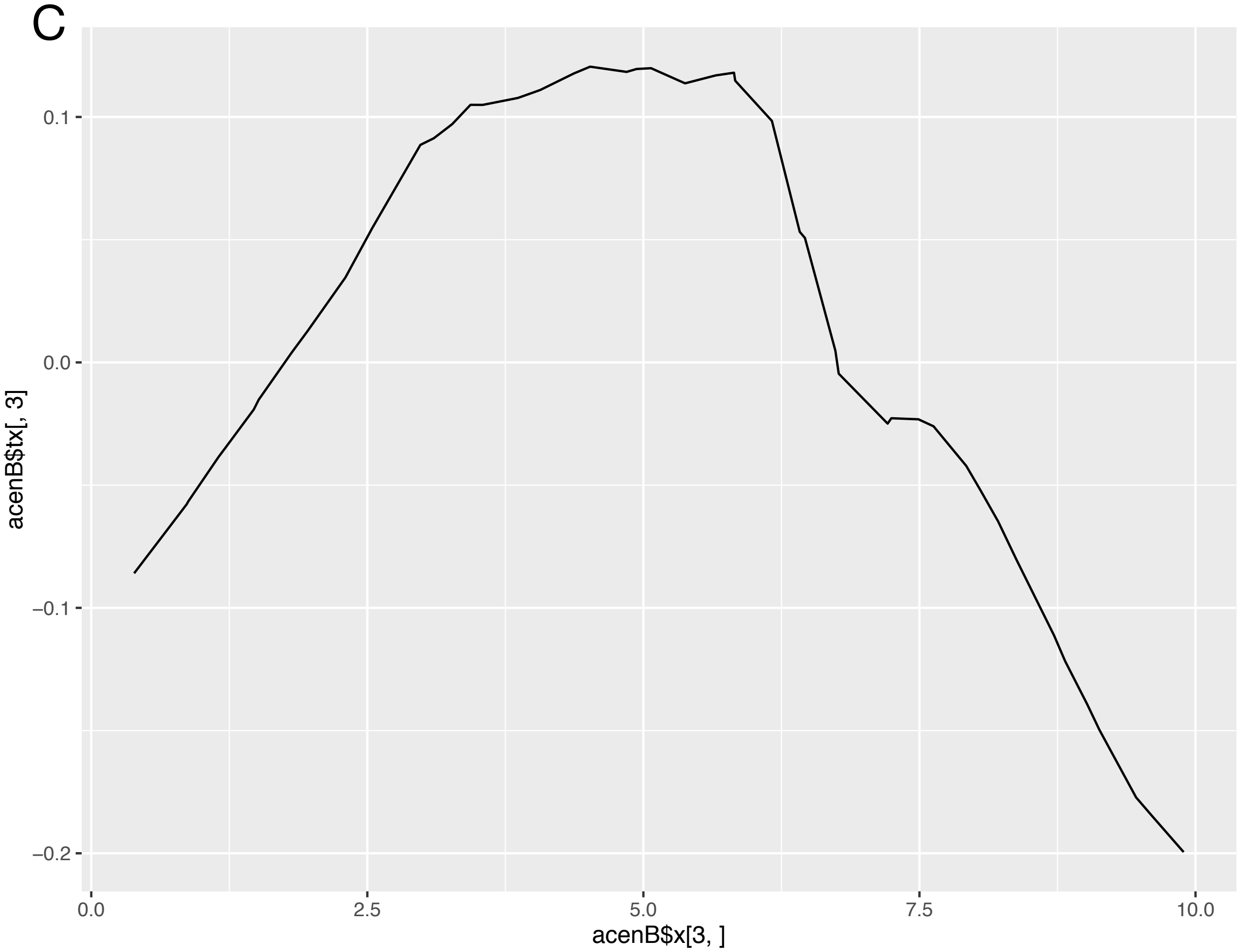

Figure 12

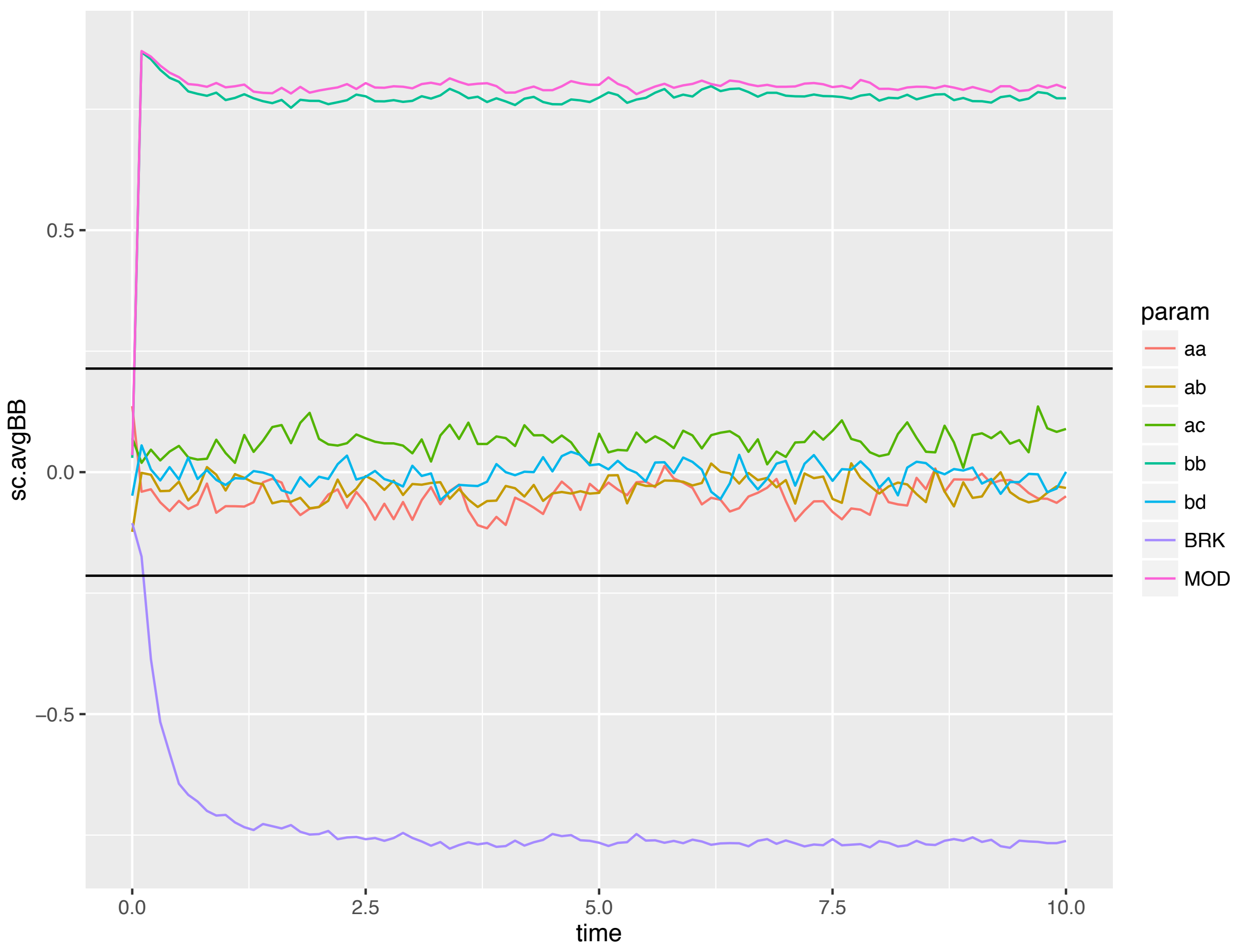

Figure 13